\documentclass[final,twoside]{raa}   
\usepackage{graphicx, times}    
\usepackage{threeparttable}
\usepackage{natbib}
\usepackage{xcolor}
\usepackage{amssymb,amsmath}
\usepackage{booktabs, caption, makecell}
\bibpunct{(}{)}{;}{a}{}{,}
\usepackage[pagebackref=true]{hyperref}

\begin{document}

\newcommand{\kms}{\text{km}/\text{s}}
\newcommand{\ergsa}{\text{erg}/\text{s}/\mathring{A}/\text{cm}^2}

\newcommand{\ai}{\mathring{A}}
\newcommand{\age}{\tau_\text{age}}
\newcommand{\feh}{[\text{Fe}/\text{H}]}
\newcommand{\logz}{\log Z/Z_\odot}
\newcommand{\ebv}{E_{\text{B}-\text{V}}}

\newcommand{\ha}{\text{H}\alpha}
\newcommand{\hb}{\text{H}\beta}

\title{Mock Observations for the CSST Mission: Integral Field Spectrograph -- \textsc{GEHONG}: A Package for Generating Ideal Datacubes}

   \volnopage{Vol.0 (20xx) No.0, 000--000}      
   \setcounter{page}{1}          

   \author{Shuai Feng\inst{1,2} \and Shiyin Shen\inst{3} \and Wei Chen\inst{3} \and Zhaojun Yan\inst{3} \and Renhao Ye\inst{3} \and Jianjun Chen\inst{4} \and Xuejie Dai\inst{3} \and Junqiang Ge\inst{4} \and Lei Hao\inst{3} \and Ran Li\inst{5,6} \and Yu Liang\inst{3} \and Lin Lin\inst{3} \and Fengshan Liu\inst{4} \and Jiafeng Lu\inst{7} \and Zhengyi Shao\inst{3} Maochun Wu\inst{3} \and Yifei Xiong\inst{3} \and Chun Xu\inst{3} \and  Yang Yang\inst{3} \and Jun Yin\inst{3} 
   }

   \institute{College of Physics, Hebei Key Laboratory of Photophysics Research and Application, Hebei Normal University, Shijiazhuang 050024, China; {\it sfeng@hebtu.edu.cn}\\
    \and
        Shijiazhuang Key Laboratory of Astronomy and Space Science / Guoshoujing Institute of Astronomy, Hebei Normal University, Shijiazhuang 050024, China\\
    \and
        Shanghai Astronomical Observatory, Chinese Academy of Sciences, 80 Nandan Road, Shanghai 200030, China {\it ssy@shao.ac.cn}\\
    \and
        National Astronomical Observatories, Chinese Academy of Sciences, Chaoyang District, Beĳing 100101, China \\
    \and
        School of Physics and Astronomy, Beijing Normal University, Beijing 100875, China \\
    \and
        School of Astronomy and Space Science, University of Chinese Academy of Science, Beijing 100049, China \\
    \and
        Institute for Astronomy, School of Physics, Zhejiang University, Hangzhou 310027, China
\vs\no
   {\small Received 20xx month day; accepted 20xx month day}}

\abstract{We developed a Python package \textsc{GEHONG} to mock the three-dimensional spectral data cube under the observation of an ideal telescope for the Integral Field Spectrograph of the Chinese Space Station Survey Telescope (CSST-IFS). This package can generate one-dimensional spectra corresponding to local physical properties at specific positions according to a series of two-dimensional distributions of physical parameters of target sources. In this way, it can produce a spatially resolved spectral cube of the target source. Two-dimensional distributions of physical parameters, including surface brightness, stellar population, and line-of-sight velocity, can be modeled using the parametric model or based on real observational data and numerical simulation data. For the generation of one-dimensional spectra, we have considered four types of spectra, including the stellar continuum spectra, ionized gas emission lines, AGN spectra, and stellar spectra. That makes \textsc{GEHONG} able to mock various types of targets, including galaxies, AGNs, star clusters, and HII regions. 
}
   \authorrunning{Feng et al.}         
   \titlerunning{\textsc{GEHONG}: A Package for Mock CSST-IFS Datacubes}  

   \maketitle
%
%
\section{Introduction}\label{sect:intro}

Integral Field Spectroscopy (IFS) has become a powerful observational technique in modern astronomy. Unlike traditional slit or fiber spectroscopy, which captures spectral information at only discrete locations, IFS simultaneously obtains a full spectrum at each position within a two-dimensional field of view. This results in a three-dimensional data cube, with two spatial dimensions and one spectral dimension, enabling spatially resolved studies of extended astronomical objects, such as galaxies. Compared to conventional spectroscopic methods, IFS provides a more comprehensive view of the internal structure, kinematics, and chemical composition of galaxies, making it an essential tool for understanding galaxy formation and evolution.

In recent years, IFS observations have significantly advanced our understanding of galaxies. These observations have revealed differences in star formation histories and stellar populations among different galaxies \citep{Gonzalez-Delgado2015, Goddard2017}, demonstrated the impacts of star formation on the surrounding ionized gas \citep{Sarzi2010, Belfiore2016}, traced the chemical enrichment of galaxies \citep{Sanchez2014, Belfiore2017}, characterized the detailed kinematic information of stars and gas \citep{Cappellari2006, Genzel2011}, and uncovered the outflows driven by AGN feedback \citep{Cano-Diaz2012, Venturi2018}. The advantages of IFS make it an indispensable tool for contemporary and future astronomical observations.

Modern telescopes are now widely equipped with IFS instruments, allowing for large-scale spectroscopic surveys of galaxies. Key facilities include the Multi Unit Spectroscopic Explorer (MUSE) on the Very Large Telescope (VLT; \citealt{Bacon2010}), the Keck Cosmic Web Imager (KCWI) on the Keck Observatory (\citealt{Morrissey2018}), and the Near Infrared Spectrograph (NIRSpec) on the James Webb Space Telescope (JWST; \citealt{Boker2022}). Several major surveys have utilized IFS to explore galaxy evolution in unprecedented detail, including the Calar Alto Legacy Integral Field Area (CALIFA) survey (\citealt{Sanchez2012}), the Mapping Nearby Galaxies at Apache Point Observatory (MaNGA) survey (\citealt{Bundy2015}), and the Sydney-AAO Multi-object Integral field spectrograph (SAMI) Galaxy Survey (\citealt{Croom2012}). These projects have provided extensive datasets that have significantly advanced our understanding of galaxy evolution.

The Chinese Space Station Telescope (CSST, \citealt{Zhan2011, Zhan2021, Gong2019}) is a 2-meter space telescope planned to be launched, which shares the same orbit with the Chinese Space Station (CSS, also known as Tiangong). The Integral Field Spectrograph is also one of the key precision instruments on board the CSST (CSST-IFS), designed for spatially resolved spectral observations of selected targets. The CSST-IFS has a $6^{''}\times6^{''}$ field of view with a spatial resolution of $0.2^{''}$. It offers a spectral coverage of $3500\ai$ to $10000\ai$ with a spectral resolution of $R \sim 1000$. With such excellent spatial resolution capabilities, especially in the ultraviolet-optical bands, the CSST-IFS will be a unique instrument for studying the fine structure of galaxies at small scales, such as star-forming regions and the vicinity of supermassive black holes within the center of galaxies.

To better understand the CSST-IFS capability, we have developed a Python package specifically to mock the three-dimensional (3D) spectroscopic data cubes of its main scientific targets. The package is named \textsc{GEHONG} (GEnerate tHe data Of iNtegral field spectrograph of Galaxy), which is dedicated to generating a wide variety of high spatial resolution ($0.1^{''}\times0.1^{''}$) 3D data cubes of galactic objects while incorporating as many necessary physical processes as possible. The 3D data cube output of \textsc{GEHONG} mocks the ideal scientific data as observed by an IFS instrument on a perfect telescope, without including instrumental or observational effects such as seeing, cosmic rays, or background light. These instrumental and observational effects are incorporated by the another software of the CSST-IFS \citep{Yan2026}, which uses the idealized data cubes generated by \textsc{GEHONG} as input to produce the corresponding raw CCD images of mock observations. These mock raw CCD images are then the key input of the scientific processing system of the CSST-IFS. Not only that, the 3D data cube output of \textsc{GEHONG} is also designed to be compatible with the input of the IFS exposure time calculator (ETC). With that, ETC can be used to quickly estimate the observation mode of the specific IFS targets and the corresponding signal-to-noise ratio,  which is helpful for the scientific pre-research of the CSST-IFS. Obviously, \textsc{GEHONG}'s output can be applied not only to CSST-IFS-related research but also to other high spatial resolution IFS instruments. In fact, similar packages have been developed for specific IFS instruments, such as \textsc{SIMSPIN}\citep{Harborne2020} and \textsc{RealSim-IFS}\citep{Bottrell2022} for SAMI, MaNGA, and MUSE. However, while these tools are tailored for these instruments, \textsc{GEHONG} remains a versatile tool that can be applied to simulate data for a wide range of IFS instruments, offering flexibility for research beyond the specific configurations of these surveys.

In this paper, we introduce the design philosophy of \textsc{GEHONG}, as well as its detailed implementation process and usage scheme. The organization of this paper is as follows: In Section \ref{sec:overview}, we present the overall framework of \textsc{GEHONG}. Sections \ref{sec:spec1d}, \ref{sec:map2d}, and \ref{sec:cube3d} describe the generation of one-dimensional spectral data, two-dimensional maps, and the final three-dimensional data cube, respectively. Finally, we summarize the key points in Section \ref{sec:sum}.

\section{Overview of \text{gehong} Package}\label{sec:overview}

\begin{table}[]
\centering
\caption{Input parameters of \texttt{config}}
\begin{tabular}{lrrl}
\hline
Parameters          & Default & Units  & Description                         \\
\hline
\texttt{wave\_min}  & 3000    & $\ai$  & Blue-end of wavelength coverage     \\
\texttt{wave\_max}  & 10500   & $\ai$  & Red-end of wavelength coverage      \\
\texttt{dlam}       & 1.5     & $\ai$  & Wavelength width of each spaxel     \\
\texttt{nx}         & 100     &        & Number of pixels in the spatial dimension (x-axis direction) \\
\texttt{ny}         & 100     &        & Number of pixels in the spatial dimension (y-axis direction) \\
\texttt{dpix}       & 0.1     & arcsec & Pixel size in the spatial dimension \\
\hline
\end{tabular}
\label{tab:config}
\end{table}

\begin{figure}
    \centering
    \includegraphics[width=\textwidth]{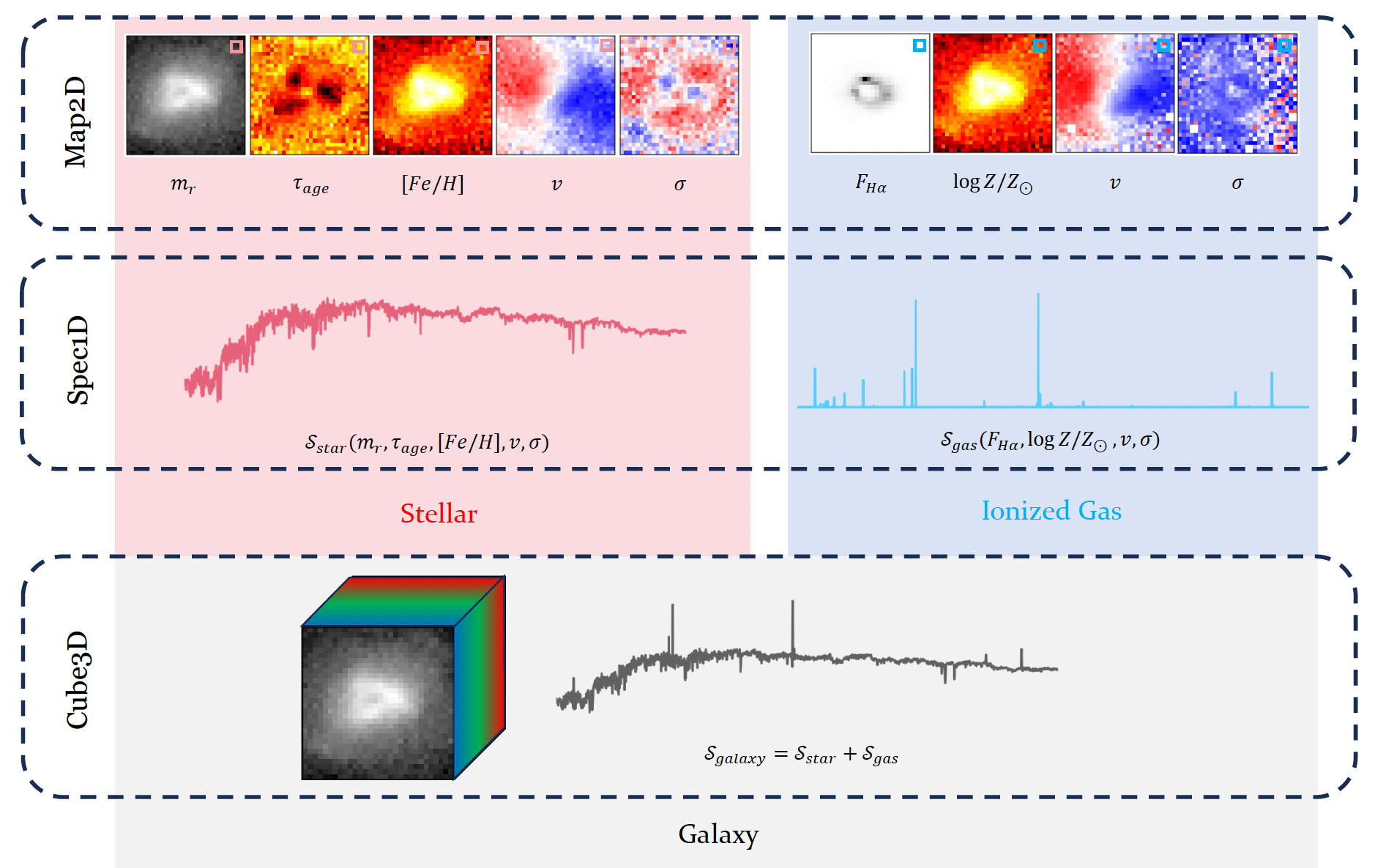}
    \caption{Schematic diagram of mocking a galaxy IFS data. }
    \label{fig:SimulatonSchema}
\end{figure}

The \textsc{GEHONG} package consists of four modules. The \texttt{config} module defines the format of the IFS data, with input parameters listed in Table \ref{tab:config}. Under the default settings, the resulting 3D data has a larger field of view, a broader wavelength range, and higher spectral and spatial resolutions compared to the CSST-IFS observation format, making it more suitable for the development of CSST-IFS’s scientific processing system. Based on these settings, the data cube $\mathcal{C}(x, y, \lambda)$ is a three-dimensional array with dimensions of $100 \times 100 \times 5000$. In addition to simulating CSST-IFS data, \textsc{GEHONG} can also generate idealized observation data for other IFS instruments by adjusting these parameters. For example, to simulate IFS data with a larger field of view, simply modify the field-of-view parameters \texttt{nx} and \texttt{ny} in the \texttt{config} module.

The generation of the IFS data cube is carried out by three key modules: \texttt{spec1d}, \texttt{map2d}, and \texttt{cube3d}, as shown in the workflow in Figure \ref{fig:SimulatonSchema}. To generate a data cube for an extended object, such as a galaxy, we first create a series of two-dimensional maps of physical parameters $\mathcal{M}(x, y)$, describing the properties at each spatial position within the object. These maps are generated by the \texttt{map2d} module (see Section \ref{sec:map2d} for details) and include stellar population properties (e.g., surface brightness, age, and metallicity), ionized gas properties (e.g., $\ha$ flux and gas-phase metallicity), and kinematics (e.g., line-of-sight velocity and velocity dispersion).

For each spatial position $(x_i, y_i)$, the corresponding one-dimensional spectrum $\mathcal{S}_i(\lambda)$ is generated using the physical parameters from the two-dimensional map $\mathcal{M}(x, y)$. This is implemented by the \texttt{spec1d} module (see Section \ref{sec:spec1d} for details), which synthesizes the light emitted from stellar populations or ionized gas components to produce the spectrum at each spatial point. The input parameters required to generate these one-dimensional spectra are provided in Table \ref{tab:spec1d}. For normal galaxies, the spectrum includes both the stellar continuum ($\mathcal{S}_\text{star}$) and the emission lines ($\mathcal{S}_\text{gas}$) from ionized gas regions.

Finally, in the \texttt{cube3d} module (see Section \ref{sec:cube3d} for details), these one-dimensional spectra are spatially arranged and combined to form the three-dimensional data cube $\mathcal{C}(x, y, \lambda)$, which represents the complete spectroscopic information across both spatial and spectral dimensions, providing a detailed view of the object’s structure, gas, and stellar properties at each point.

The above provides an introduction to the generation of IFS data for extended sources (e.g., galaxies and HII regions). For point sources (e.g., AGN, stars, and star clusters), only one-dimensional spectra need to be inserted at specific positions, eliminating the need for constructing two-dimensional maps. Since nearby galaxies are the primary targets of CSST-IFS, the following sections will focus on galaxies as examples to illustrate the technical details of \textsc{GEHONG}. For instructions on using \textsc{GEHONG}, please refer to the appendix.

\section{One-Dimensional Spectrum}\label{sec:spec1d}

\begin{table}[]
\caption{Input Parameters of \texttt{spec1d}}
\label{tab:spec1d}
\begin{threeparttable}
\centering
\begin{tabular}{llrl}
\hline
Parameters            & Units                  & Example\tnote{1} & Description                                       \\ \hline
\multicolumn{4}{c}{Stellar Population Continuum (\texttt{spec1d.StellarContinuum})} \\ \hline
\texttt{mag}          & mag                    & $15.0$      & Magnitude in SDSS-$r$ band \\
\texttt{sfh}          & Gyr                    & $1.0$       & Star formation history or single age (Gyr) \\
\texttt{ceh}          & dex                    & $-0.3$     & Chemical enrichment history or single metallicity ([Fe/H]) \\
\texttt{vel}          & $\kms$                 & $100.0$    & Line-of-sight velocity of stellar continuum \\
\texttt{vdisp}        & $\kms$                 & $100.0$    & Line-of-sight velocity dispersion of stellar continuum \\
\texttt{ebv}          & mag                    & $0.1$      & Dust reddening ($\ebv$) of stellar continuum \\ \hline
\multicolumn{4}{c}{Ionized Gas Emission Line (\texttt{spec1d.HII\_Region})} \\ \hline
\texttt{halpha}       & $10^{-17}$erg/s/cm$^2$ & $200$     & Integral flux of $\ha$ emission line            \\
\texttt{logz}         & dex                    & $-0.2$    & Gas-phase metallicity ($\logz$)                 \\
\texttt{vel}          & $\kms$                 & $30000$   & Line-of-sight velocity of ionized gas \\
\texttt{vdisp}        & $\kms$                 & $150$     & Line-of-sight velocity dispersion of ionized gas \\
\texttt{ebv}          & mag                    & $0.1$     & Dust reddening ($\ebv$) of ionized gas \\ \hline
\multicolumn{4}{c}{Single Star Spectrum (\texttt{spec1d.SingleStar})} \\ \hline
\texttt{mag}          & mag                    & $15$      & Magnitude in SDSS-$r$ band                      \\
\texttt{teff}         & K                      & $8000$    & Effective temperature                           \\
\texttt{feh}          & dex                    & $-0.1$    & Metallicity ($\feh$) of single stellar          \\
\texttt{logg}         & cm/s$^2$               & $3$       & Surface gravity ($\log g$) of single stellar          \\
\texttt{vel}          & $\kms$                 & $800$     & Line-of-sight velocity of single stellar   \\
\texttt{ebv}          & mag                    & $0.1$     & Dust reddenning ($\ebv$) of single stellar \\ \hline
\multicolumn{4}{c}{AGN Spectrum}                                                                             \\ \hline
\multicolumn{4}{c}{Power Law Continuum (\texttt{spec1d.AGN\_Powerlaw})}  \\
\texttt{m5100}        & mag                    & $17$      & Magnitude between $5050\ai$ and $5150\ai$ at the restframe \\
\texttt{alpha}        &                        & $-1.5$    & Spectrum index of power law                     \\ 
\texttt{vel}          & $\kms$                 & $30000$   & Line-of-sight velocity of AGN                         \\
\texttt{ebv}          & mag                    & $0.1$     & Dust reddenning ($\ebv$) of AGN                      \\ \hline
\multicolumn{4}{c}{Broad Emission Lines (\texttt{spec1d.AGN\_BLR})} \\
\texttt{hbeta\_flux} & $10^{-17}$erg/s/cm$^2$  & $800$      & Integral flux of $\hb$ broad line              \\
\texttt{hbeta\_fwhm}  & $\kms$                 & $5000$    & FWHM of $\hb$ broad line                        \\ 
\texttt{vel}          & $\kms$                 & $30000$   & Line-of-sight velocity of AGN                        \\
\texttt{ebv}          & mag                    & $0.1$     & Dust reddenning ($\ebv$) of AGN                      \\ \hline
\multicolumn{4}{c}{Narrow Emission Lines (\texttt{spec1d.AGN\_NLR})} \\
\texttt{halpha}       & $10^{-17}$erg/s/cm$^2$ & $500$     & Integral flux of $\ha$ narrow line              \\
\texttt{logz}         & dex                    & $-0.3$    & Gas-phase metallicity ($\logz$)                 \\
\texttt{vdisp}        & $\kms$                 & $800$     & Line-of-sight velocity dispersion of narrow emission lines            \\ 
\texttt{vel}          & $\kms$                 & $30000$   & Line-of-sight velocity of AGN                         \\
\texttt{ebv}          & mag                    & $0.1$     & Dust reddenning ($\ebv$) of AGN                      \\ \hline
\multicolumn{4}{c}{Fe II Emission Lines (\texttt{spec1d.AGN\_FeII})} \\
\texttt{r4570}        &                        & $0.4$     & Flux ratio between Fe4570 and $\hb$ broad line  \\ 
\texttt{hbeta\_broad} & $10^{-17}$erg/s/cm$^2$ & $800$     & Integral flux of $\hb$ broad line               \\ 
\texttt{vel}          & $\kms$                 & $30000$   & Line-of-sight velocity of AGN                        \\
\texttt{ebv}          & mag                    & $0.1$     & Dust reddening ($\ebv$) of AGN                   \\ \hline
\end{tabular}
\begin{tablenotes}
    \item[1] Input parameters of the mock spectrum examples in Figures 2, 3, and 4.
\end{tablenotes}
\end{threeparttable}
\end{table}

\subsection{Continuum of Stellar Population}\label{sec:spec1d_ssp}

\begin{figure*}
    \centering
    \includegraphics[width=\linewidth]{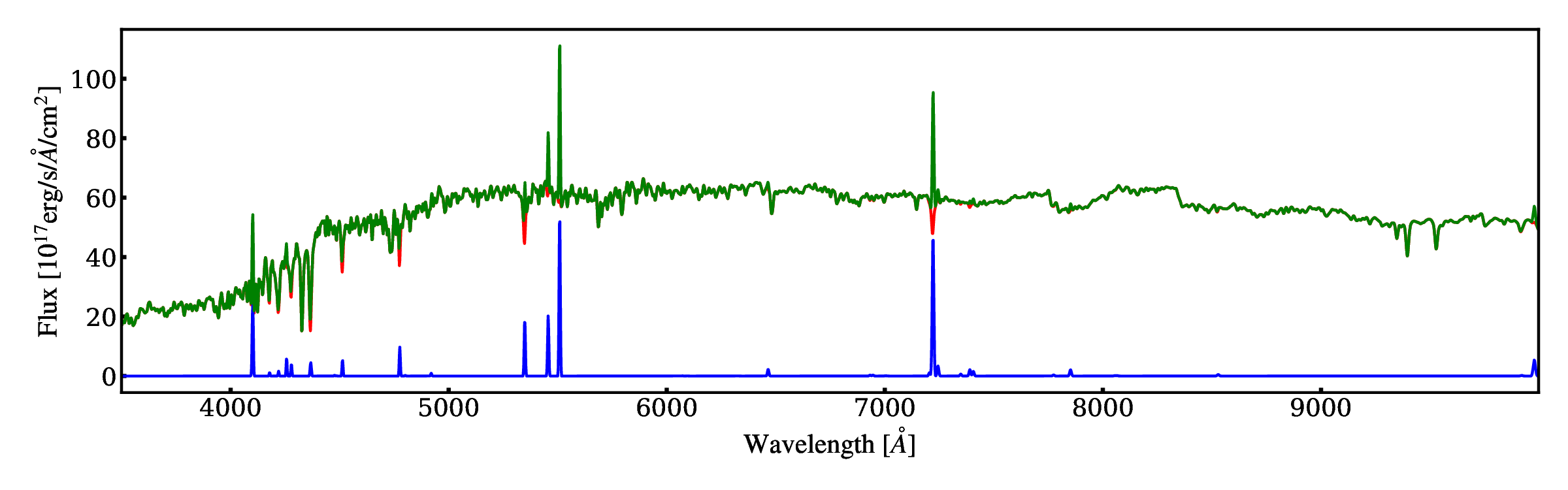}
    \caption{An example of a synthesized spectrum of a normal galaxy. The green line shows the integrated spectrum of the galaxy, the red line represents the stellar population continuum, and the blue line represents the ionized gas emission lines. The spectrum is generated using the \texttt{spec1d} and \texttt{map2d} modules, with the input parameters summarized in Table~\ref{tab:spec1d} and Table~\ref{tab:map2d}. A simple example demonstrating the generation of the galaxy spectrum is provided in Appendix~\ref{app:spec1d}.}
    \label{fig:ExampleStellarPop}
\end{figure*}

The continuum of the stellar population is generated by the \texttt{spec1d.StellarContinuum} module. An example spectrum is shown as the red solid line in Figure~\ref{fig:ExampleStellarPop}, with the corresponding input parameters listed in Table~\ref{tab:spec1d}. The continuum is constructed using single stellar population (SSP) template spectra as the fundamental building blocks. In this work, we adopt the E-MILES stellar population models\footnote{\url{http://miles.iac.es/}} \citep{Vazdekis2016}, which provide a broad wavelength coverage from $1680\ \text{\AA}$ to $50000\ \text{\AA}$, fully encompassing the sensitivity range of CSST-IFS. The E-MILES library also offers extensive sampling in stellar age and metallicity, enabling flexible modeling of a wide variety of galaxy types.

In \textsc{GEHONG}, the stellar continuum for each spatial pixel is assembled by combining SSP templates according to a user-specified star formation history (SFH) and chemical enrichment history (CEH). The SFH and CEH can be supplied as two-dimensional arrays, where the first column specifies lookback time and the second column provides either the relative star formation rate (SFR) or the stellar metallicity ([Fe/H]) at each epoch. Only the relative distribution of star formation over time is used to determine the template weights; absolute normalization is not considered. The resulting composite stellar population (CSP) spectrum is produced by linearly summing the weighted SSP templates. Alternatively, for fast and simplified applications, users can input a single age and metallicity. In this case, the continuum is generated by directly selecting the SSP template whose age and metallicity are closest to the input values. This dual-mode approach enables both detailed and efficient modeling depending on scientific requirements.

Starting from the assembled stellar continuum, several physical effects are subsequently applied to generate the final output spectrum.

First, we account for the broadening of the stellar continuum caused by stellar velocity dispersion. The E-MILES templates already incorporate spectral broadening due to the finite instrumental resolution of the original observations, with a wavelength-dependent resolution that corresponds to an intrinsic dispersion of approximately $200\ \text{km}\ \text{s}^{-1}$ in the UV ($\lambda < 3541\ \text{\AA}$) and typically less than $100\ \text{km}\ \text{s}^{-1}$ in the optical and near-infrared. To simulate the internal velocity dispersion of galaxies, we apply additional broadening only when the input dispersion exceeds the intrinsic template dispersion. In such cases, Gaussian convolution is performed using the fast Fourier method in logarithmic wavelength space, following the implementation in the \texttt{PPXF} package \citep{Cappellari2017}. Second, internal dust attenuation is applied based on the input reddening value $\ebv$, following the attenuation law of \citet{Calzetti2000} and assuming no foreground reddening from the Milky Way. The reddened spectrum is computed as 
\begin{equation}\label{eq:reddening}
\mathcal{S}(\lambda) = \mathcal{S}_\text{nodust}(\lambda) \times e^{0.921 \ebv k(\lambda)}, 
\end{equation}
where $k(\lambda)$ denotes the attenuation curve. Third, the spectrum is shifted to the observer’s frame by applying a redshift correction using the input line-of-sight velocity ($v_\text{stellar}$), with the observed wavelength given by 
\begin{equation}\label{eq:redshift} 
\lambda_\text{obs} = \lambda_\text{rest} \left(1 + \frac{v_\text{stellar}}{c}\right), 
\end{equation} 
where $c$ is the speed of light. Finally, the spectrum is flux-calibrated to match the specified apparent magnitude in the SDSS-$r$ band. The flux is rescaled accordingly, and the final output is expressed in units of $10^{-17}\ \text{erg}\ \text{s}^{-1}\ \text{\AA}^{-1}\ \text{cm}^{-2}$.

The input parameters for the \texttt{spec1d.StellarContinuum} module is summarized in Table~\ref{tab:spec1d}, and a simple guide to its usage is provided in Appendix~\ref{app:spec1d}.

\subsection{Emission Lines of Ionized Gas}\label{sec:emline}

In this work, the emission lines of ionized gas are modeled considering only the contribution from HII regions, implemented through the \texttt{spec1d.HII\_Region} module. An example spectrum of ionized gas emission is shown as the green solid line in Figure~\ref{fig:ExampleStellarPop}.

The emission lines are generated by representing each line as a Gaussian profile, with the width reflecting the velocity dispersion of the ionized gas. We include $84$ emission lines spanning the wavelength range from $900\ \text{\AA}$ to $10500\ \text{\AA}$. Each emission line is indexed by $i$, with a central wavelength $\lambda_i$. The profile of the $i$-th emission line is given by 
\begin{equation} 
\mathcal{E}_i(\lambda) = \frac{1}{\sigma_{\text{line},i} \sqrt{2\pi}} \exp\left( -\frac{(\lambda - \lambda_i)^2}{2\sigma_{\text{line},i}^2} \right), 
\end{equation} 
where $\sigma_{\text{line},i}$ is the Gaussian width (in \text{\AA}) of the $i$-th line. Within each spaxel, a constant velocity dispersion $\sigma_\text{gas}$ is assumed for all emission lines. The width $\sigma_{\text{line},i}$ is derived from the ionized gas velocity dispersion $\sigma_\text{gas}$ (in km~s$^{-1}$) according to 
\begin{equation} 
\sigma_{\text{line},i} = \frac{\sigma_\text{gas}}{c} \lambda_i, 
\end{equation} 
where $c$ is the speed of light. The composite emission line spectrum is constructed by summing over all individual lines: 
\begin{equation} 
\mathcal{S}(\lambda) = \sum_{i=1}^{N} \mathcal{L}_i \mathcal{E}_i(\lambda), 
\end{equation} 
where $\mathcal{L}_i$ is the relative flux of the $i$-th emission line normalized to the H$\alpha$ flux.

The relative fluxes of the emission lines are determined based on the emission line models from \citet{Byler2017}, who used \texttt{Cloudy} \citep{Ferland2013} simulations to model HII regions ionized by young stellar clusters. In these models, the line ratios depend on gas-phase metallicity, the cluster age, and the ionization parameter. For simplicity, we vary only the metallicity, while fixing the cluster age at $10^6$~yr and the ionization parameter at $\log U = -2$.\footnote{The ionized gas emission lines and stellar continuum are treated independently. The cluster age used here characterizes typical HII regions and is unrelated to the stellar population ages.}

Dust reddening and redshift correction are applied based on the properties of the ionized gas. Dust reddening is applied using the reddening parameter of ionized gas following Equation~\ref{eq:reddening}, and redshift correction is performed based on the gas line-of-sight velocity following Equation~\ref{eq:redshift}. The final emission line spectrum is then flux-scaled to match a given integrated H$\alpha$ flux, expressed in units of $10^{-17}\ \text{erg}\ \text{s}^{-1}\ \text{cm}^{-2}$.

The list of input parameters for ionized gas emission modeling is provided in Table~\ref{tab:spec1d}, and an illustrative usage example is presented in Appendix~\ref{app:spec1d}.

\subsection{Spectrum of Active Galactic Nuclei}\label{sec:spec1d_agn}

\begin{figure}
    \centering
    \includegraphics[width=\textwidth]{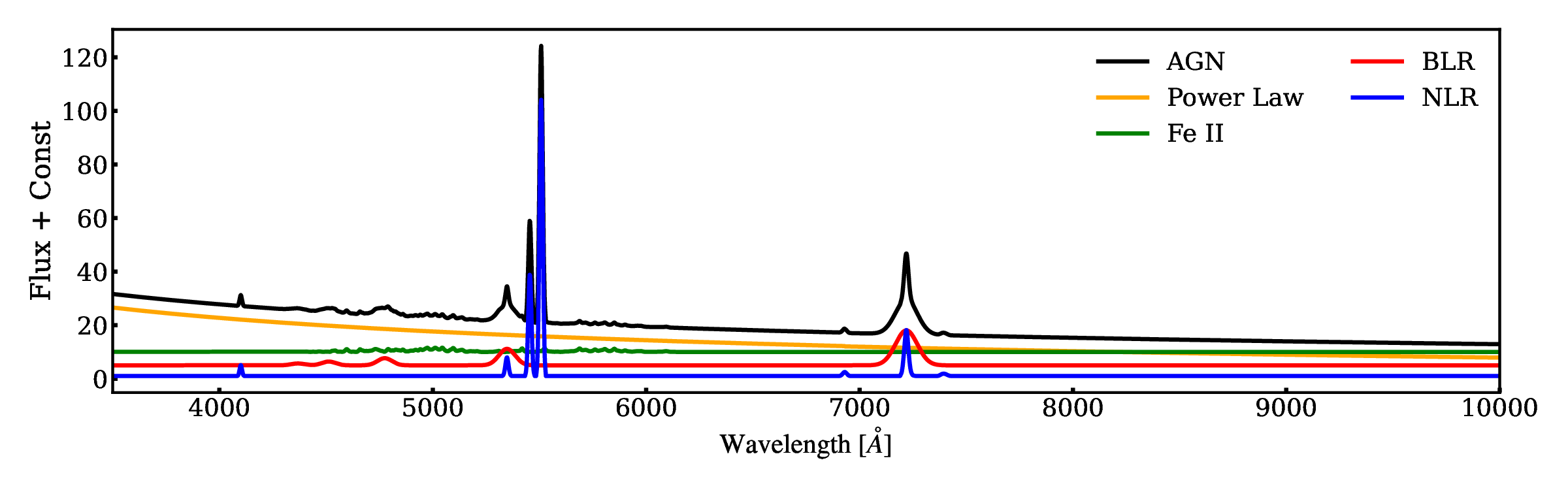}
    \caption{An example of a generated AGN spectrum (black line), decomposed into its four components: the power-law continuum (yellow line), the iron emission line spectrum (green line), the broad-line region (BLR) emission lines (red line), and the narrow-line region (NLR) emission lines (blue line). The input parameters for each component are summarized in Table~\ref{tab:spec1d}. A simple example demonstrating the generation of the AGN spectrum is provided in Appendix~\ref{app:spec1d}.}
    \label{fig:AGN_Spec}
\end{figure}

Typically, the spectra of AGNs (Active Galactic Nuclei) consist of four main components: a power-law continuum, a broad-line region (BLR) emission line spectrum, a narrow-line region (NLR) emission line spectrum, and an iron emission line spectrum. In our framework, the mock spectra of AGNs are generated by separately simulating these four components and then combining them. The power-law continuum, BLR spectrum, NLR spectrum, and iron emission spectrum are generated using the modules \texttt{spec1d.AGN\_Powerlaw}, \texttt{spec1d.AGN\_BLR}, \texttt{spec1d.AGN\_NLR}, and \texttt{spec1d.AGN\_FeII}, respectively. The input parameters for each component are summarized in Table~\ref{tab:spec1d}, and a simple example illustrating the usage of these modules is provided in Appendix~\ref{app:spec1d}.

In the following subsections, we describe the modeling logic for each of the four spectral components. For each component, dust reddening and redshift effects are individually applied following the same procedure as described for the stellar continuum in Section~\ref{sec:spec1d_ssp}. Since the treatment is identical across all components, we do not repeat these details in each subsection.

\subsubsection{Power-law Spectrum}\label{sec:spec1d_powerlaw}

The power-law continuum component of AGNs (orange line in Figure~\ref{fig:AGN_Spec}) is generated using the \texttt{spec1d.AGN\_Powerlaw} module. The spectrum is modeled as 
\begin{equation}
    F(\lambda) = F_{5100} \times \left( \frac{\lambda}{5100\ \text{\AA}} \right)^{-\alpha}, 
\end{equation} 
where $F_{5100}$ is the flux density at $5100\ \text{\AA}$ in the rest frame, and $\alpha$ is the spectral index. The overall flux normalization is determined by $F_{5100}$, which sets the absolute scaling of the spectrum. In practice, $F_{5100}$ is derived from the apparent magnitude at $5100\ \text{\AA}$ in the rest frame.

\subsubsection{Narrow-Line Region Spectrum}\label{sec:spec1d_nlr}

The narrow-line region (NLR) spectrum of AGNs (blue line in Figure~\ref{fig:AGN_Spec}) is generated using the \texttt{spec1d.AGN\_NLR} module. The generation method follows the approach described in Section~\ref{sec:emline}, with the difference that an AGN-specific narrow-line region model is adopted instead of the HII region model. We employ the model developed by \citet{Feltre2016}, which includes ten strong optical emission lines. In this model, the line intensity ratios depend on several physical parameters, including the gas-phase metallicity ($\log Z/Z_\odot$), ionization parameter ($\log U$), metal-to-dust ratio ($\xi_\text{d}$), neutral hydrogen density ($\log n_\text{H}/\text{cm}^{-3}$), and UV photon spectral index ($\alpha$). In our framework, only the gas-phase metallicity is treated as a free parameter, while the other parameters are fixed at $\log U = -1$, $\xi_\text{d} = 0.1$, $\log n_\text{H}/\text{cm}^{-3} = -2$, and $\alpha = -1.4$. The absolute flux normalization of the NLR spectrum is set by the integrated flux of the narrow component of the H$\alpha$ emission line, and the line widths are determined by the velocity dispersion of the same narrow H$\alpha$ component. The treatment of flux scaling and line broadening follows the same procedure as described for HII regions in Section~\ref{sec:emline}.

\subsubsection{Broad-Line Region Spectrum}\label{sec:spec1d_blr}

The broad-line region (BLR) spectrum of AGNs (red line in Figure~\ref{fig:AGN_Spec}) is generated using the \texttt{spec1d.AGN\_BLR} module. The BLR spectrum is modeled as a collection of broad emission lines, following a methodology similar to that described in Section~\ref{sec:emline}. In contrast to the emission lines from HII regions, only the broad components of the five Balmer lines ($\text{H}\epsilon$, $\text{H}\delta$, $\text{H}\gamma$, $\hb$, and $\ha$) are included. The relative fluxes among these lines are fixed based on the observational measurements of Mrk~817 \citep{Ilic2006}. The line widths are set according to the full width at half maximum (FWHM) of the broad $\hb$ emission line, following the common observational convention. The absolute flux normalization of the BLR spectrum is determined by the integrated flux of the broad $\hb$ component.

\subsubsection{Iron Emission Line Spectrum}\label{sec:spec1d_ironline}

The iron emission line spectrum of AGNs (green line in Figure~\ref{fig:AGN_Spec}) is generated using the \texttt{spec1d.AGN\_FeII} module. The spectral shape is modeled based on the empirical Fe~\textsc{II} emission line template provided by \citet{Park2022}, which covers the wavelength range from $4000\ \text{\AA}$ to $5600\ \text{\AA}$ with a resolution of approximately $2\ \text{\AA}$. The absolute flux normalization of the Fe~\textsc{II} spectrum is determined by the flux ratio parameter $R4570 = \text{Fe}\textsc{II}\lambda4570 / \hb$, where $\text{Fe}\textsc{II}\lambda4570$ refers to the integrated flux of the Fe~\textsc{II} blend centered at $4570\ \text{\AA}$, and $\hb$ denotes the broad $\hb$ emission line \citep{Marinello2016}. Given a specified $R4570$ and the integrated flux of the broad $\hb$ component, the absolute flux of the Fe~\textsc{II} spectrum is anchored to the flux scale of the BLR emission.

\subsection{Single Stellar Spectrum}\label{sec:spec1d_singlestellar}

\begin{figure*}
    \centering
    \includegraphics[width=\linewidth]{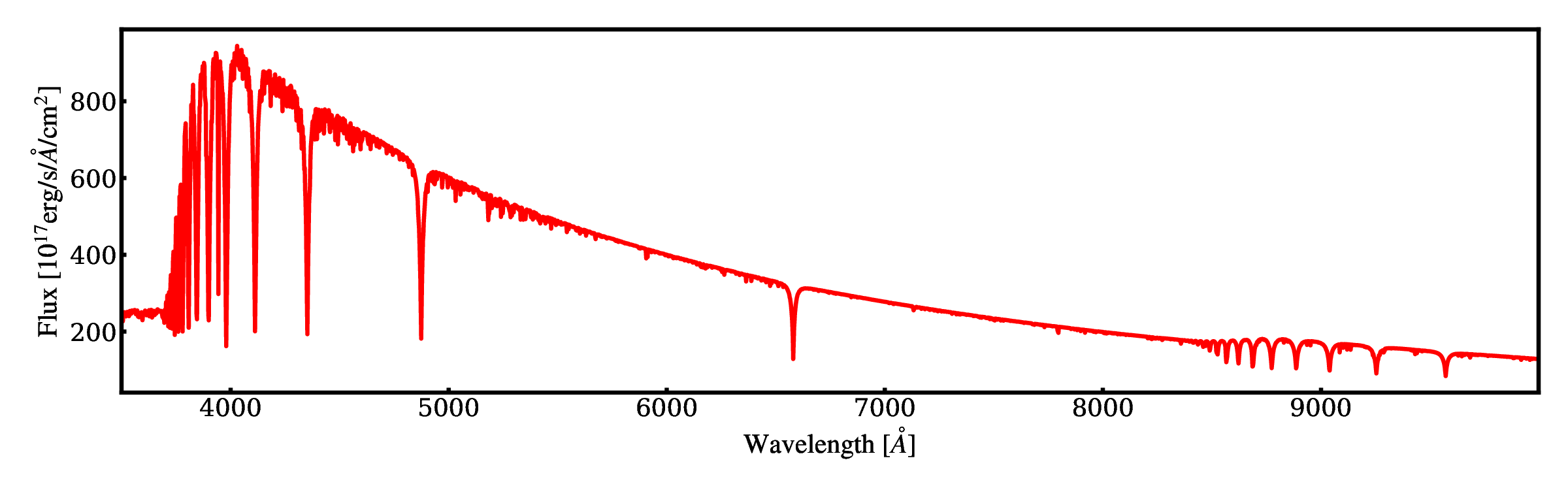}
    \caption{An example of a synthesized spectrum of a single star. The spectrum is generated using the \texttt{spec1d.SingleStar} module, with the input parameters summarized in Table~\ref{tab:spec1d}. A simple example demonstrating the generation of the stellar spectrum is provided in Appendix~\ref{app:spec1d}.}
    \label{fig:StarSpec1D}
\end{figure*}

The spectra of single stars are generated using the \texttt{spec1d.SingleStar} module. In Figure~\ref{fig:StarSpec1D}, we present an example of a single stellar spectrum, shown by the red solid line. The input parameters used to generate this spectrum are summarized in Table~\ref{tab:spec1d}, and a simple example demonstrating the generation process is provided in Appendix~\ref{app:spec1d}.

This module employs the stellar spectral templates from \citet{Munari2005}, which cover a wavelength range from $2500\ \text{\AA}$ to $10500\ \text{\AA}$ and are available at multiple spectral resolutions. We adopt the set with a uniform dispersion of $1\ \text{\AA}/\text{pixel}$. The templates span the full Hertzsprung-Russell diagram, with effective temperatures ranging from $3500,\text{K} \leq T_\text{eff} \leq 47500,\text{K}$, surface gravities from $0.0 \leq \log g \leq 5.0$, and metallicities from $-2.5 \leq \text{[Fe/H]} \leq 0.5$. In addition, variations in $\alpha$-abundance and rotational velocity are also considered in the templates.

In our current implementation, the shape of the stellar spectrum is determined by specifying three physical parameters: effective temperature, metallicity, and surface gravity. Other parameters are fixed to typical values, assuming zero rotational velocity and solar $\alpha$-abundance. The appropriate template is selected by first matching the input metallicity and surface gravity, followed by choosing the template with the closest effective temperature.

After selecting the optimal template, the effects of dust reddening and redshift are applied using Equation~\ref{eq:reddening} and Equation~\ref{eq:redshift}, based on the specified dust reddening and line-of-sight velocity. Finally, the absolute flux calibration is performed by matching the SDSS $r$-band apparent magnitude, ensuring the spectrum is normalized appropriately for observational applications.

\section{Two-Dimensional Maps} \label{sec:map2d}

\begin{table}
\centering
\caption{Input Parameters of \text{map2d}}
\label{tab:map2d}
\begin{threeparttable}
\begin{tabular}{llrl}
\hline
Parameter & Units  & Example\tnote{2} & Description                          \\ \hline
\multicolumn{4}{c}{Sersic Model (\texttt{map2d.sersic\_map})} \\ \hline
\texttt{mag}       & mag    & 15      & Integral magnitude of Sersic model   \\
\texttt{reff}      & arcsec & 4       & Effective radius                     \\
\texttt{n}         &        & 1.0     & Sersic index                         \\
\texttt{ellip}     &        & 0.6     & Ellipticity                          \\
\texttt{pa}        & degree & 30      & Position angle                       \\ \hline
\multicolumn{4}{c}{tahn Model (\texttt{map2d.tanh\_map})} \\ \hline
\texttt{vmax}      & km/s   & 160     & Maximum rotational velocity          \\
\texttt{rt}        & arcsec & 2       & Turn-over radius of rotation curve   \\
\texttt{ellip}     &        & 0.8     & Ellipticity                          \\
\texttt{pa}        & degree & 0       & Position angle                       \\ \hline
\multicolumn{4}{c}{Gredient Model (\texttt{map2d.gred\_map})} \\ \hline
\texttt{aeff}\tnote{2} &    & 9.5     & Amplitude at the effective radius    \\
\texttt{reff}      & arcsec & 6       & Effective radius                     \\
\texttt{gred}      &        & -1.2    & Gredient                             \\
\texttt{ellip}     &        & 0.4     & Ellipticity                          \\
\texttt{pa}        & degree & 45      & Position angle                       \\ \hline
\end{tabular}
\begin{tablenotes}
    \item[1] Input parameters of the mock map examples in Figure 5.
    \item[2] For the case of the age map, the units of \texttt{aeff} and \texttt{gred} are $\log \text{yr}$. 
\end{tablenotes}
\end{threeparttable}
\end{table}

To generate mock IFS data for extended sources such as galaxies, it is essential to first construct a series of two-dimensional maps of physical parameters. These 2D maps provide the necessary input for each spaxel, supplying quantities such as stellar population properties and ionized gas emission strengths, which are subsequently used by the \texttt{spec1d} module to generate individual one-dimensional spectra. The accuracy and spatial structure of these maps directly determine the realism of the resulting mock IFS data.

Taking galaxies as an example, the 2D parameter maps can be broadly divided into two categories. The first category describes the stellar component, including quantities required for producing stellar continuum spectra as discussed in Section~\ref{sec:spec1d_ssp}, such as the surface brightness distribution, the stellar age distribution, and the stellar metallicity distribution. These maps are assembled using the \texttt{map2d.StellarPopulationMap} module, which accepts a set of two-dimensional arrays as input. The second category pertains to the gaseous component, encompassing parameters necessary for modeling ionized gas emission lines as described in Section~\ref{sec:emline}, such as the spatial distribution of H$\alpha$ flux and the gas-phase metallicity. These maps are organized through the \texttt{map2d.IonizedGasMap} module, also based on two-dimensional arrays.

To ensure that the 2D parameter maps closely resemble the complex structures observed in real galaxies, we recommend constructing them based on high-spatial-resolution observational data, such as imaging from the Hubble Space Telescope (HST) or integral field spectroscopy from MUSE. From such observations, key physical maps—including surface brightness distributions, stellar population property maps, and kinematic maps—can be extracted and used to define the spatial variation of input parameters. Alternatively, outputs from cosmological simulations, such as the IllustrisTNG project, can also be utilized to generate detailed two-dimensional distributions. When direct observational or simulation data are unavailable, \textsc{GEHONG} provides several commonly used parametric models to facilitate the construction of 2D maps. In the following subsections, we describe three representative parametric models implemented in \textsc{GEHONG}.

\begin{figure}
    \centering
    \includegraphics[width = \textwidth]{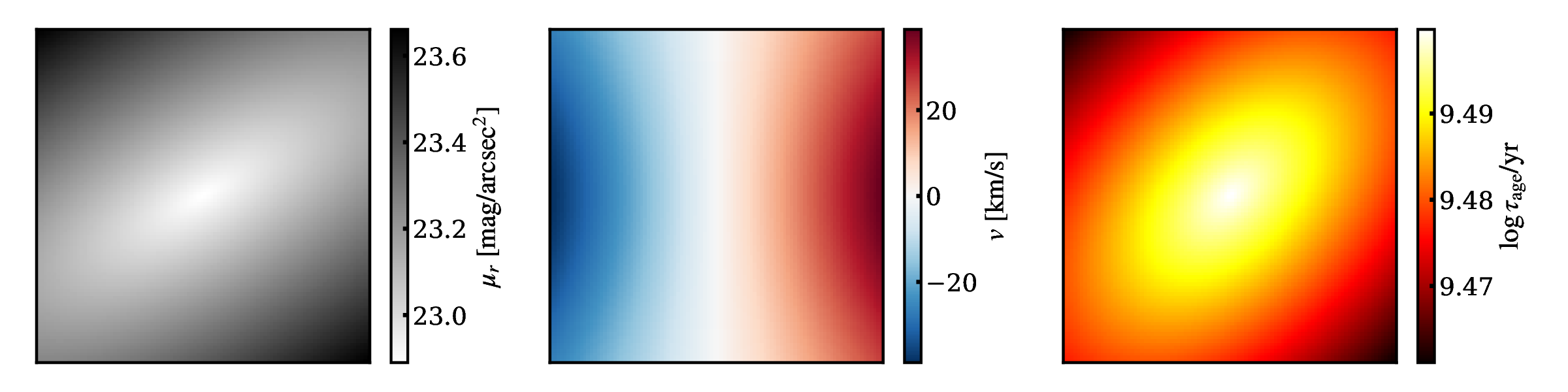}
    \caption{An example of two-dimensional maps of physical parameters generated using the \texttt{map2d} module, including surface brightness, line-of-sight velocity, and stellar population age. The input parameters used to create the maps are summarized in Table~\ref{tab:map2d}, and a simple example demonstrating the generation of the maps is provided in Appendix~\ref{app:map2d}.}
    \label{fig:case_map2d}
\end{figure}

\subsection{\texttt{map2d.sersic\_map}}

To model the surface brightness distribution of galaxies, \textsc{GEHONG} provides the \texttt{map2d.sersic\_map} module, which generates a two-dimensional Sérsic profile \citep{Sersic1963, Graham2005}. This parametric model can effectively describe a wide range of galaxy light distributions, from exponential disks to de Vaucouleurs bulges. An example of a Sérsic-based surface brightness map is shown in the left panel of Figure~\ref{fig:case_map2d}, with the corresponding input parameters listed in Table~\ref{tab:map2d}. A simple code example demonstrating the usage of this module is provided in Appendix~\ref{app:map2d}.

The Sérsic profile is described by the following equation: 
\begin{equation} 
    I(x, y) = I_e \exp\left \{-b_n\left [\left (\frac {R(x, y)}{R_e}\right)^{1/n}-1\right]\right \}
\end{equation} 
where $I_e$ is the surface brightness at the effective radius $R_e$, $n$ is the Sérsic index that controls the concentration of the profile, and $b_n$ is a constant depending on $n$, determined by 
\begin{equation} 
    \Gamma(2n) = 2\gamma(2n, b_n). 
\end{equation} 
Here, $\Gamma$ and $\gamma$ denote the complete and incomplete gamma functions, respectively.

The effective radius $R_e$ represents the half-light radius, and the Sérsic index $n$ characterizes the profile shape: $n=1$ corresponds to an exponential profile typical of disk galaxies, while $n=4$ corresponds to a de Vaucouleurs profile typical of elliptical galaxies.

The radial distance $R(x,y)$ from the galaxy center $(x_0, y_0)$ to a point $(x, y)$ accounts for the galaxy inclination and ellipticity, and is given by 
\begin{equation}\label{eq:Radius} 
    R(x, y) = \sqrt{R_\text{maj}^2(x, y) + \left( \frac{R_\text{min}(x, y)}{1-q} \right)^2}, 
\end{equation} 
where $q$ is the minor-to-major axis ratio (related to ellipticity) and $\theta$ is the position angle. The coordinates along the major and minor axes are defined as 
\begin{align} 
    R_\text{maj}(x, y) &= (x - x_0)\cos\theta + (y - y_0)\sin\theta, \\ 
    R_\text{min}(x, y) &= -(x - x_0)\sin\theta + (y - y_0)\cos\theta. 
\end{align}

To calibrate the surface brightness normalization, the total apparent magnitude $m_\text{tot}$ is used. The corresponding total luminosity $L_\text{tot}$ for a galaxy following a Sérsic profile can be expressed as \citep{Ciotti1991}: 
\begin{equation} 
    L_\text{tot} = 2\pi \int_0^{\infty} I(R) R, \mathrm{d}R = I_e R_e^2 2\pi n \frac{e^{b_n}}{(b_n)^{2n}} \Gamma(2n). 
\end{equation}

\subsection{\texttt{map2d.tanh\_map}}

For modeling the velocity field of rotating galaxies, \textsc{GEHONG} implements the \texttt{map2d.tanh\_map} module. This module generates an axisymmetric rotation map based on a hyperbolic tangent rotation curve \citep{Andersen2013}, a common approximation for disk galaxies. The middle panel of Figure~\ref{fig:case_map2d} shows an example of such a velocity map, with the input parameters summarized in Table~\ref{tab:map2d}. A code example for generating this type of velocity field is provided in Appendix~\ref{app:map2d}.

For galaxies with an inclination angle approximately given by $i \approx \arccos(1-q)$, where $q$ is the axis ratio, the line-of-sight velocity at a position $(x, y)$ on the galaxy plane is expressed as \citep{vanderKruit1978}: 
\begin{equation} 
    V(x,y) = V_\text{sys} + V_c(R)\cos\phi\sin I, 
\end{equation} 
where $V_\text{sys}$ is the systemic recession velocity, $V_c(R)$ is the intrinsic rotational velocity at radius $R$, and $\phi$ is the azimuthal angle in the galaxy plane.

The spatial coordinates $(x, y)$ are related to $(R, \phi)$ through the expressions given in Equation~\ref{eq:Radius}. The azimuthal angle $\phi$ is determined by: 
\begin{equation} 
    \cos \phi = \frac{-(x-x_0)\sin\theta + (y-y_0)\cos\theta}{R(x,y)}, 
\end{equation} 
where $(x_0, y_0)$ is the galaxy center and $\theta$ is the position angle.

The intrinsic rotational velocity $V_c(R)$ is modeled by a hyperbolic tangent function: 
\begin{equation} 
    V_c(R) = V_\text{max} \tanh\left(\frac{R}{R_t}\right), 
\end{equation} 
where $V_\text{max}$ is the maximum rotational velocity, and $R_t$ is the turnover radius at which the rotation curve flattens \citep{Andersen2013}.

\subsection{\texttt{map2d.gred\_map}}

In addition to surface brightness and velocity, many physical parameters such as stellar age, stellar metallicity, and gas-phase metallicity exhibit approximately radial gradients in galaxies \citep{Koleva2011, SanchezBlazquez2014, Belfiore2017}. \textsc{GEHONG} provides the \texttt{map2d.gred\_map} module to construct such parameter maps. An example of a radial gradient map is shown in the right panel of Figure~\ref{fig:case_map2d}, and the corresponding input parameters are listed in Table~\ref{tab:map2d}. The usage of this module is illustrated with a code example in Appendix~\ref{app:map2d}.

For a galaxy with an inclination angle $i$, a physical parameter $A$ at a position $(x, y)$ is described by: 
\begin{equation} 
    A(x,y) = A(R, \theta) = A_\text{eff} + \nabla_A \log\left(\frac{R(x, y)}{R_e}\right), 
\end{equation} 
where $A_\text{eff}$ is the value of $A$ at the effective radius $R_e$, and $\nabla_A$ represents the logarithmic gradient of $A$ with respect to radius. The radial distance $R(x, y)$ and the azimuthal angle $\theta$ are defined as in Equation~\ref{eq:Radius}.

\section{Three-Dimensional Cube}\label{sec:cube3d}

As mentioned in Section~\ref{sec:overview} and illustrated in Figure~\ref{fig:SimulatonSchema}, the primary task of three-dimensional datacube generation is to arrange and integrate one-dimensional spectra according to the spatial positions provided by the two-dimensional parameter maps. In practice, this is accomplished using the \texttt{cube3d} module.

\subsection{Extended Source}

For extended sources such as galaxies, the mock datacube is constructed by generating a one-dimensional spectrum at each spatial pixel based on the local physical parameters. Specifically, given a two-dimensional map $\mathcal{M}(x, y)$, at each position $(x_i, y_i)$, we mock a one-dimensional spectrum $\mathcal{S}_i(\lambda)$ according to $\mathcal{M}(x_i, y_i)$. This spectrum is then assigned to $\mathcal{C}(x_i, y_i, \lambda)$ in the three-dimensional datacube. By repeating this process over all spatial pixels, we obtain the full mock datacube $\mathcal{C}(x, y, \lambda)$. The input to the \texttt{cube3d} module consists of the classes generated by the \texttt{map2d} module.

The spectrum of a galaxy typically consists of both stellar continuum and ionized gas emission lines. Accordingly, the two-dimensional maps used for the mock of the datacube should include information on both stellar populations (e.g., stellar population age and metallicity) and ionized gas properties (e.g., $\ha$ emission line flux and gas-phase metallicity). If the mocked galaxy does not exhibit significant emission lines, as is typical for early-type galaxies, only stellar population maps are required. Conversely, when mocking pure emission-line sources, such as HII regions, only ionized gas maps need to be provided.

The above procedure applies to target sources with relatively simple structures. For more complex cases, it is necessary to decompose the target into several simpler components. The datacube of each component is mocked separately and then combined to obtain the final datacube of the target source. For example, to mock a galaxy exhibiting strong ionized gas outflows, the system should be divided into at least two parts: a main galaxy and an outflowing ionized gas component. The main galaxy datacube, including both the stellar continuum and the normal ionized gas emission lines, is mocked following the method described in the previous paragraph. The outflow component is treated as a pure emission-line source, requiring only the mock of ionized gas spectra. After generating the individual mock datacubes for the main galaxy and the outflow, they are combined to produce the final datacube representing the galaxy with strong gas outflows.

\subsection{Point Source}

In addition to extended sources, \textsc{GEHONG} also supports the mock of point sources, such as AGNs or individual stars. For point sources, the mock datacube is constructed by assigning a one-dimensional spectrum to a specific spatial pixel, without any spatial extension. In the current implementation, no point spread function (PSF) convolution is applied during this assignment, and the surrounding spaxels remain empty.

The \textsc{GEHONG} framework is designed to mock datacubes as they would be observed by an idealized telescope, assuming perfect optics without PSF blurring, detector noise, or other instrumental effects. It focuses on modeling the intrinsic spatial and spectral properties of target sources. The simulation of realistic observational effects, such as replicating the actual performance of CSST-IFS, is handled by another dedicated software (Yan et al., in preparation), which incorporates instrumental and observational effects.

Among various observational effects, the PSF is particularly important for point sources, as it redistributes the flux across multiple adjacent spaxels, leading to spatial broadening and dilution of the central intensity. In future versions of \textsc{GEHONG}, we plan to implement a simplified PSF convolution option, such as applying a Gaussian kernel, to enable approximate modeling of spatial blurring effects when needed.

\section{Summary}\label{sec:sum}

Integral Field Spectroscopy (IFS) has revolutionized the study of galaxies by providing three-dimensional data cubes that simultaneously capture spectral and spatial information. The upcoming Chinese Space Station Telescope (CSST) will be equipped with a high-spatial-resolution Integral Field Spectrograph (CSST-IFS) to enable detailed investigations of the internal structures of galaxies. To support scientific preparation and instrument optimization for CSST-IFS, we have developed \textsc{GEHONG}, a Python package designed to mock IFS datacubes for various astrophysical targets under idealized observational conditions.

The \textsc{GEHONG} package adopts a modular architecture for constructing synthetic IFS data, consisting of three main modules. The \texttt{spec1d} module generates one-dimensional spectra at individual spatial positions based on input physical parameters, including apparent magnitude, stellar population age and metallicity, emission line fluxes, gas-phase metallicity, line-of-sight velocity, and velocity dispersion. It supports the modeling of stellar population continua, ionized gas emission lines, Active Galactic Nucleus (AGN) spectra, and single-star spectra. The \texttt{map2d} module constructs two-dimensional distributions of physical parameters, either through parametric models—such as the Sérsic model, rotating disk model, and gradient model—or from user-defined inputs. Finally, the \texttt{cube3d} module assembles the three-dimensional data cube by assigning one-dimensional spectra to spatial pixels, enabling the mock of both extended sources, such as galaxies, and compact sources such as AGNs.

\begin{acknowledgements}

The authors thank the anonymous referee for their constructive comments and suggestions, which helped improve the quality of this manuscript. This work is supported by the CSST scientific data processing and analysis system of the China Manned Space Project. S.F. acknowledges support from National Natural Science Foundation of China (Nos. 12103017), Natural Science Foundation of Hebei Province (No. A2021205001, A2025205037). S.S. thanks research grants from the Shanghai Academic/Technology Research Leader (22XD1404200), the National Key R\&D Program of China (No. 2022YFF0503402 ), National Natural Science Foundation of China (No. 12141302), and the China Manned Space Project with No. CMS-CSST-2025-A07. J.G. acknowledges support from the National Astronomical Observatories of the Chinese Academy of Sciences (No. E4ZR0510), the Beijing Municipal Natural Science Foundation (No. 1242032), the National Key Research and Development Program of China (No. 2023YFA1607904), and the Youth Innovation Promotion Association of the Chinese Academy of Sciences (No. 2022056).

\end{acknowledgements}

\appendix

\section{Usage of \textsc{GEHONG}}\label{app:usage}

The \textsc{GEHONG} package is publicly available at \url{https://csst-ifs-gehong.readthedocs.io/}. Users can install the package via \texttt{pip} using the following command: 
\begin{verbatim} 
    pip install csst-ifs-gehong 
\end{verbatim} 

Comprehensive documentation and detailed usage instructions are provided on the website. For the convenience of readers, we also present the example codes used to generate the mock cases below, as shown in the figures throughout this paper. The following code example is based on version 3.1.0 of \textsc{GEHONG}.

\subsection{\texttt{spec1d} Module}\label{app:spec1d}

This section provides example codes for generating the mock spectra presented in the figures throughout the paper using the \texttt{spec1d} module.

To reproduce the ionized gas emission lines shown in Figure~\ref{fig:ExampleStellarPop}, the following codes can be used: 
\begin{verbatim} 
    # Set the configuration of data format 
    conf = config.config() 
    # Load the HII region emission line template 
    gas_tem = spec1d.EmissionLineTemplate(conf, model='hii') 
    # Generate the ionized gas emission lines of an HII region 
    gas = spec1d.HII_Region(conf, gas_tem, halpha=500, logz=-0.2, 
    vel=30000, vdisp=150, ebv=0.1) 
\end{verbatim}

The continuum of the stellar population, also shown in Figure~\ref{fig:ExampleStellarPop}, can be generated as follows: 
\begin{verbatim} 
    # Load the stellar population templates 
    stellar_tem = spec1d.StellarContinuumTemplate(conf) 
    # Generate the stellar population continuum 
    stellar = spec1d.StellarContinuum(conf, stellar_tem, mag=17, sfh=2, 
    ceh=-0.3, vel=30000, vdisp=150, ebv=0.1) 
\end{verbatim}

The four components of the AGN spectrum illustrated in Figure~\ref{fig:AGN_Spec} can be mocked separately using the following codes: 
\begin{verbatim} 
    # Load the emission line template for the narrow-line region 
    nlr_temp = spec1d.EmissionLineTemplate(conf, model='nlr') 
    # Generate the power-law continuum spectrum 
    pl = spec1d.AGN_Powerlaw(conf, m5100=18, alpha=-1.5, vel=30000, 
    ebv=0.1) 
    # Generate the iron emission line spectrum 
    fe = spec1d.AGN_FeII(conf, hbeta_broad=800.0, r4570=0.4, 
    ebv=0.1, vel=30000) 
    # Generate the broad-line region spectrum 
    blr = spec1d.AGN_BLR(conf, hbeta_flux=800, hbeta_fwhm=5000.0, 
    vel=30000, ebv=0.1) 
    # Generate the narrow-line region spectrum 
    nlr = spec1d.AGN_NLR(conf, nlr_temp, halpha=500, logz=-0.3, 
    vel=30000, vdisp=400, ebv=0.1) 
\end{verbatim}

Finally, the spectrum of a single star shown in Figure~\ref{fig:StarSpec1D} can be generated using: 
\begin{verbatim} 
    # Load the template for a single star 
    star_temp = spec1d.SingleStarTemplate(conf) 
    # Generate the single stellar spectrum 
    star = spec1d.SingleStar(conf, star_temp, mag=15, teff=8000, 
    feh=-0.1, logg=3, vel=800, ebv=0.1) 
\end{verbatim}

\subsection{\texttt{map2d} Module}\label{app:map2d}

The two-dimensional maps in Figure \ref{fig:case_map2d} are mocked using the following codes:
\begin{verbatim}
    # Initialize the surface brightness map
    sbmap = map2d.Map2d(config)
    # Mocking the map of surface brightness
    sbmap.sersic_map(mag=15.0, reff=4, n=1.0, ellip=0.6, theta=30)
    # Initialize the velocity map
    velmap = map2d.Map2d(config)
    # Mocking the map of velocity
    velmap.tanh_map(vmax=160, rt=2.0, ellip=0.8, theta=0)
    # Initialize the stellar age map
    agemap = map2d.Map2d(config)
    # Mocking the map of stellar age
    agemap.gred_map(aeff=9.5, reff=6, gred=-0.2, ellip=0.4, theta=45)
\end{verbatim}

\bibliographystyle{raa}
\bibliography{bibtex}

\label{lastpage}

\end{document}